\documentclass[conference]{IEEEtran}
\IEEEoverridecommandlockouts
\usepackage{cite}
\usepackage{amsmath,amssymb,amsfonts}
\usepackage{algorithmic}
\usepackage{graphicx}
\usepackage{textcomp}
\usepackage{xcolor}

\usepackage{booktabs}
\usepackage[linesnumbered,ruled,vlined]{algorithm2e}
\usepackage{colortbl}
\usepackage{caption}
\usepackage{multirow}

\definecolor{tabfirst}{rgb}{1, 0.7, 0.7} % red
\definecolor{tabsecond}{rgb}{1, 0.85, 0.7} % orange
\definecolor{tabthird}{rgb}{1, 1, 0.7} % yellow

\def\BibTeX{{\rm B\kern-.05em{\sc i\kern-.025em b}\kern-.08em
    T\kern-.1667em\lower.7ex\hbox{E}\kern-.125emX}}

\begin{document}
% \twocolumn[{%
% \renewcommand\twocolumn[1][]{#1}%
% \maketitle

% \centering
% \includegraphics[width=\linewidth]{images/teaser2.pdf}
% \captionof{figure}{Our custom-designed backpack-mounted mobile system enables the synchronized capture of images and LiDAR point clouds for reconstruction in textureless and geometrically degenerate real-world scenes. (a) Our method on data captured with the custom MLS device. (b) 3DGS results on SfM-based data.}
% \label{fig:teaser}
% }]

\title{GTLR-GS: Geometry-Texture Aware LiDAR-Regularized 3D Gaussian Splatting for Realistic Scene Reconstruction
% \thanks{Identify applicable funding agency here. If none, delete this.}
}

% \author{\IEEEauthorblockN{Anonymous Authors}}
\author{
\IEEEauthorblockN{Yan Fang\IEEEauthorrefmark{1}, Jianfei Ge\IEEEauthorrefmark{1} Jiangjian Xiao\IEEEauthorrefmark{1}}
\IEEEauthorblockA{\IEEEauthorrefmark{1}
Ningbo Institute of Materials Technology and Engineering, UCAS\\
Ningbo, China\\
Email: fangyan23@mails.ucas.ac.cn}
}

% \author{\IEEEauthorblockN{1\textsuperscript{st} Given Name Surname}
% \IEEEauthorblockA{\textit{dept. name of organization (of Aff.)} \\
% \textit{name of organization (of Aff.)}\\
% City, Country \\
% email address or ORCID}
% \and
% \IEEEauthorblockN{2\textsuperscript{nd} Given Name Surname}
% \IEEEauthorblockA{\textit{dept. name of organization (of Aff.)} \\
% \textit{name of organization (of Aff.)}\\
% City, Country \\
% email address or ORCID}
% \and
% \IEEEauthorblockN{3\textsuperscript{rd} Given Name Surname}
% \IEEEauthorblockA{\textit{dept. name of organization (of Aff.)} \\
% \textit{name of organization (of Aff.)}\\
% City, Country \\
% email address or ORCID}
% \and
% \IEEEauthorblockN{4\textsuperscript{th} Given Name Surname}
% \IEEEauthorblockA{\textit{dept. name of organization (of Aff.)} \\
% \textit{name of organization (of Aff.)}\\
% City, Country \\
% email address or ORCID}
% \and
% \IEEEauthorblockN{5\textsuperscript{th} Given Name Surname}
% \IEEEauthorblockA{\textit{dept. name of organization (of Aff.)} \\
% \textit{name of organization (of Aff.)}\\
% City, Country \\
% email address or ORCID}
% \and
% \IEEEauthorblockN{6\textsuperscript{th} Given Name Surname}
% \IEEEauthorblockA{\textit{dept. name of organization (of Aff.)} \\
% \textit{name of organization (of Aff.)}\\
% City, Country \\
% email address or ORCID}
% }

\maketitle

\begin{abstract}
Recent advances in 3D Gaussian Splatting (3DGS) have enabled real-time, photorealistic scene reconstruction. However, conventional 3DGS frameworks typically rely on sparse point clouds derived from Structure-from-Motion (SfM), which inherently suffer from scale ambiguity, limited geometric consistency, and strong view dependency due to the lack of geometric priors. In this work, a LiDAR-centric 3D Gaussian Splatting framework is proposed that explicitly incorporates metric geometric priors into the entire Gaussian optimization process. Instead of treating LiDAR data as a passive initialization source, 3DGS optimization is reformulated as a geometry-conditioned allocation and refinement problem under a fixed representational budget. Specifically, this work introduces (i) a geometry-texture-aware allocation strategy that selectively assigns Gaussian primitives to regions with high structural or appearance complexity, (ii) a curvature-adaptive refinement mechanism that dynamically guides Gaussian splitting toward geometrically complex areas during training, and (iii) a confidence-aware metric depth regularization that anchors the reconstructed geometry to absolute scale using LiDAR measurements while maintaining optimization stability. Extensive experiments on the ScanNet++ dataset and a custom real-world dataset validate the proposed approach. The results demonstrate state-of-the-art performance in metric-scale reconstruction with high geometric fidelity.

\end{abstract}


\begin{figure*}[t]
    \centering
    \includegraphics[width=\linewidth]{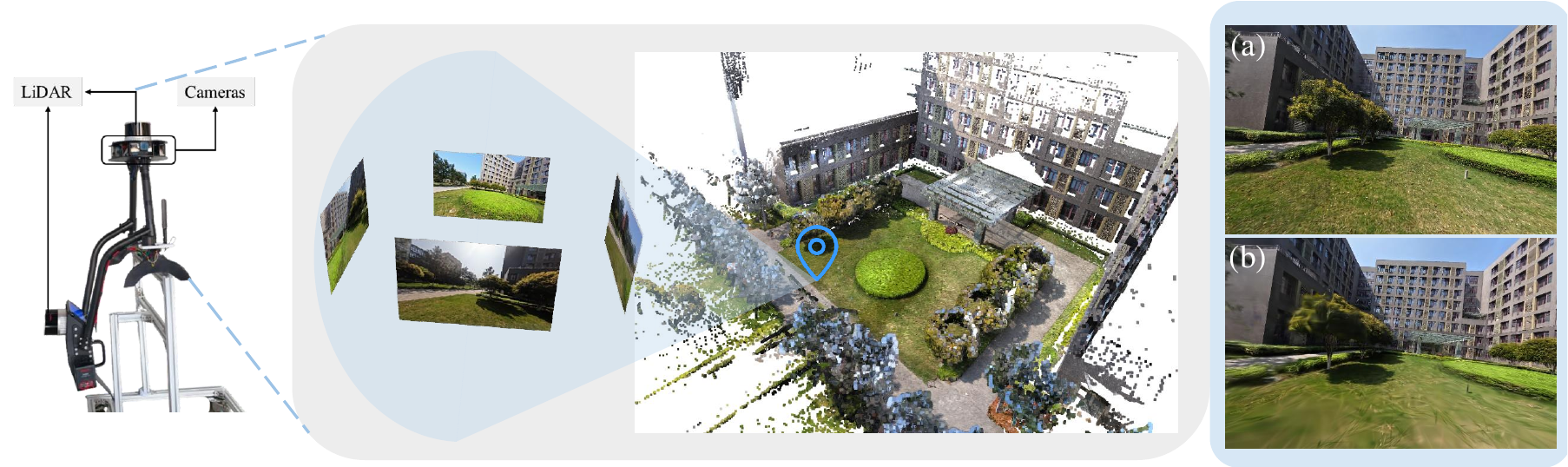}
    \caption{Our custom-designed backpack-mounted mobile system enables the synchronized capture of images and LiDAR point clouds for reconstruction in textureless and geometrically degenerate real-world scenes. (a) Our method on data captured with the custom MLS device. (b) 3DGS results on SfM-based data.}
    \label{fig:teaser}
\end{figure*}

\section{Introduction}

High-fidelity and metrically accurate 3D scene reconstruction is a fundamental requirement for a wide range of real-world applications, including industrial inspection~\cite{yang2020review}, infrastructure modeling~\cite{cao2022towards}, digital twins~\cite{liu2025citygo}, and embodied robotics~\cite{zhu20243d}. In these scenarios, reconstruction systems are required not only to produce visually plausible renderings, but also to preserve geometric structures with reliable metric scale. Achieving this goal remains challenging in practice, particularly in scenes dominated by textureless surfaces, repetitive patterns, and limited viewpoints.

Recent advances in 3D Gaussian Splatting (3DGS)~\cite{kerbl20233d} have significantly improved the efficiency of neural scene representations, enabling real-time rendering with competitive visual quality. Despite these advantages, existing 3DGS pipelines are typically initialized from sparse Structure-from-Motion (SfM) point clouds~\cite{schonberger2016structure}, which inherently suffer from scale ambiguity, incomplete geometry, and strong view dependency. As a consequence, reconstructed scenes often exhibit floating artifacts, structural distortions, and unstable geometry under novel viewpoints. Although replacing SfM initialization with dense LiDAR point clouds alleviates scale ambiguity, this substitution alone is insufficient: even with metrically accurate inputs, the standard 3DGS optimization process may still degrade fine geometric structures during training.

Analysis of this behavior reveals three key limitations that hinder the deployment of 3DGS in real-world, metric-critical settings. First, \emph{inefficient Gaussian allocation}: dense LiDAR point clouds often contain tens of millions of points, far exceeding the computational budget of 3DGS. Naive downsampling strategies, such as uniform or random sampling, ignore the highly non-uniform distribution of geometric and appearance information, leading to oversimplification in structurally complex regions and redundant allocation in planar areas. Second, \emph{geometry-unaware densification}: the adaptive density control mechanism in vanilla 3DGS relies primarily on view-dependent photometric signals and lacks explicit geometric guidance. As a result, Gaussian splitting may blur sharp edges and thin structures, even when initialized from high-quality LiDAR geometry. Third, \emph{missing metric-scale constraints}: image-based supervision alone is insufficient to prevent scale drift or suppress floating artifacts, particularly under sparse-view or texture-degenerate conditions.

These limitations indicate that LiDAR data should not be treated merely as a denser replacement for SfM initialization, but rather as a source of \emph{metric geometric priors} that actively guide Gaussian allocation, refinement, and constraint enforcement throughout training. Motivated by this insight, this work introduces \textbf{GTLR-GS}, a LiDAR-centric 3D Gaussian Splatting framework that explicitly incorporates geometric priors into all stages of the optimization process. Instead of uniformly distributing Gaussian primitives, their placement and evolution are conditioned on local geometric and appearance complexity, while metric consistency is enforced through LiDAR-derived depth supervision.

The proposed framework decomposes Gaussian optimization into three geometry-conditioned stages. First, a \emph{geometry-texture-aware adaptive sampling} strategy allocates a fixed Gaussian budget preferentially to regions with high curvature and significant texture variation, enabling efficient utilization of dense LiDAR point clouds without sacrificing structural detail. Second, a \emph{curvature-adaptive splitting} mechanism modulates Gaussian densification based on local surface complexity, preventing structural degradation commonly observed in view-dependent splitting schemes. Third, a \emph{confidence-aware metric LiDAR depth regularization} injects absolute scale constraints into the training objective, stabilizing geometry and suppressing floating artifacts under sparse or challenging viewpoints.

To evaluate performance under realistic conditions, a backpack-mounted mobile LiDAR scanning system is developed to synchronize LiDAR, IMU, and multi-camera inputs, enabling robust data acquisition in textureless and geometrically degenerate environments where SfM-based pipelines often fail. Extensive experiments on the ScanNet++ benchmark~\cite{yeshwanth2023scannet++} and a custom real-world dataset demonstrate improved rendering quality, enhanced geometric fidelity, and more efficient Gaussian utilization compared to state-of-the-art 3DGS variants.

In summary, the main contributions of this work are:
\begin{itemize}
    \item A formulation of 3D Gaussian Splatting as a geometry-conditioned allocation and refinement problem, together with a LiDAR-centric framework that explicitly leverages metric geometric priors.
    \item A geometry-texture-aware adaptive sampling strategy that allocates Gaussian primitives to information-rich regions under a fixed representational budget.
    \item A curvature-adaptive splitting mechanism that preserves fine geometric structures during densification by aligning Gaussian refinement with surface complexity.
    \item A confidence-aware LiDAR depth regularization scheme that enforces metric-scale consistency and suppresses floating artifacts in challenging real-world scenes.
\end{itemize}

\begin{figure*}[h]
	\centering
	\includegraphics[width=\linewidth]{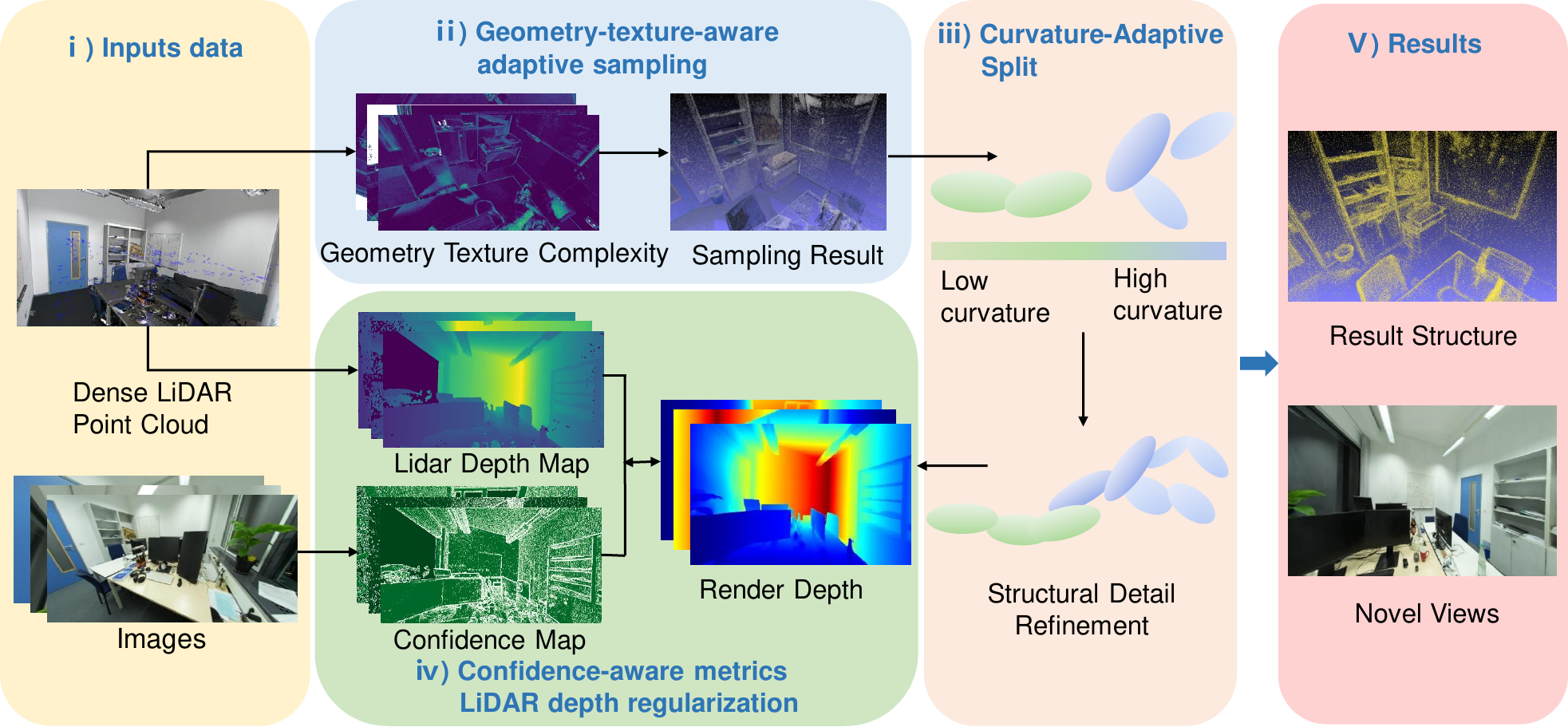}
	\caption{\textbf{GTLR-GS Overview}. \romannumeral1) The input data consists of dense LiDAR point clouds and RGB images. \romannumeral2) Geometric and texture complexity is computed to guide point cloud sampling for 3DGS initialization. \romannumeral3) Dynamically adjusts the splitting intensity based on local surface complexity. \romannumeral4) Depth maps are extracted from the registered point cloud map and used as regularization constraints to supervise 3DGS training. \romannumeral5) The trained 3DGS results include the Gaussian sphere distribution structure and novel view synthesis performance.}
	\label{fig:pipeline}
\end{figure*}

\section{Related Work}

\subsection{Gaussian Splatting with Explicit Geometry Priors}

3D Gaussian Splatting~\cite{kerbl20233d} introduces an explicit point-based scene representation that enables real-time rendering while maintaining competitive visual quality. However, its original formulation relies predominantly on photometric supervision and implicitly assumes that Gaussian primitives naturally conform to scene geometry. In practice, this assumption often breaks down, leading to floating artifacts, blurred edges, and view-dependent structural distortions.

To address these limitations, a growing body of work incorporates explicit geometric priors into the Gaussian optimization process. DN-Splatter~\cite{turkulainen2025dn} integrates monocular depth and normal predictions to regularize Gaussian placement, following a strategy similar to MonoSDF~\cite{yu2022monosdf}. DNGaussian~\cite{li2024dngaussian} further introduces hierarchical depth regularization to alleviate scale ambiguity in depth supervision, while PGSR~\cite{chen2024pgsr} proposes unbiased depth rendering with planar Gaussian primitives, enforcing geometric consistency through joint depth and normal constraints across views.

Beyond depth and normal-based regularization, hybrid frameworks combine Gaussian splatting with implicit surface representations. GSDF~\cite{yu2024gsdf} and GaussianRoom~\cite{xiang2025gaussianroom} co-optimize 3D Gaussians with neural signed distance fields, leveraging SDF gradients to guide splat distributions toward surface-consistent configurations. Related efforts such as SuGaR~\cite{guedon2024sugar} reconstruct surfaces via Poisson reconstruction from Gaussian-sampled point clouds, though they may suffer from fragmented geometry due to unstructured density fields. Structured Gaussian parameterizations, including 2DGS~\cite{huang20242d}, GOF~\cite{yu2024gaussian}, and RaDe-GS~\cite{rade-gs}, reinterpret Gaussian primitives as surface-aligned elements or rasterized splats to improve surface coherence and depth accuracy.

\subsection{LiDAR-Guided Gaussian Splatting}

With the increasing availability of high-resolution LiDAR sensors, recent works have explored incorporating LiDAR data into Gaussian splatting frameworks. One representative line of research focuses on LiDAR-GS methods designed for autonomous driving and sensor simulation. These approaches aim to reproduce realistic LiDAR observations and sensor characteristics, including ray attenuation, intensity modeling, and range noise, to support perception and simulation tasks in dynamic driving environments~\cite{chen2024lidar, jiang2025gs, zhou2025lidar}. In such settings, LiDAR is primarily used to model sensor behavior rather than to enforce geometric consistency for high-fidelity scene reconstruction.

Another line of work leverages LiDAR point clouds as geometric priors to improve reconstruction accuracy and scale consistency. For example, several methods replace or augment SfM-based initialization with LiDAR-derived point clouds to obtain metric-scale geometry~\cite{cui2025streetsurfgs, hess2025splatad, yan2024street}. However, these approaches typically treat LiDAR as a dense initialization or auxiliary supervision signal, without explicitly addressing how Gaussian primitives should be allocated, refined, and densified under a fixed computational budget. As a result, even with metric-scale inputs, Gaussian distributions may remain inefficiently allocated or suffer from structural degradation during optimization.

\subsection{Efficiency-Oriented and Sparse-View Gaussian Splatting}

Another line of research focuses on improving the efficiency and scalability of Gaussian splatting. Octree-GS~\cite{ren2025octree} introduces a hierarchical level-of-detail representation to reduce memory consumption and accelerate rendering, while FSGS~\cite{zhu2024fsgs} targets sparse-view scenarios by guiding Gaussian unpooling through proximity-based heuristics. These methods emphasize computational efficiency and scalability, often prioritizing memory and speed over fine-grained geometric control.

Recent works also explore adaptive sampling and region-based rendering strategies to reallocate computational resources toward informative views or spatial regions~\cite{wang2024adr, lin2024vastgaussian, liu2024citygaussian, cui2024letsgo, kerbl2024hierarchical}. While such approaches improve rendering quality under constrained budgets, they are typically driven by image-level or screen-space signals and lack access to explicit metric geometric priors.

\section{Method}

\subsection{Problem Formulation and Overview}

Given a set of calibrated RGB images $\mathcal{I}$ and a registered LiDAR point cloud $\mathcal{P}$ with metric-scale accuracy, our goal is to reconstruct a 3D Gaussian-based scene representation that supports high-quality novel view synthesis while preserving geometric fidelity under a limited computational budget. Following the 3DGS formulation, the scene is represented as a collection of Gaussian primitives $\mathcal{G}=\{G_i\}$, each parameterized by position, covariance, opacity, and appearance attributes. Under a fixed representational budget, effectively leveraging dense metric geometry while maintaining stable optimization remains non-trivial.

To address this challenge, 3DGS optimization is formulated as a \emph{geometry-conditioned allocation and refinement} process. Rather than treating LiDAR as a passive initialization source, metric geometric priors are explicitly used to guide Gaussian placement, refinement, and scale constraint enforcement throughout training. As illustrated in Fig.~\ref{fig:pipeline}, the proposed framework consists of three stages: geometry-conditioned allocation, geometry-aware refinement, and metric-scale constraint enforcement.

\begin{figure}[tbp]
	\centering
	\includegraphics[width=\linewidth]{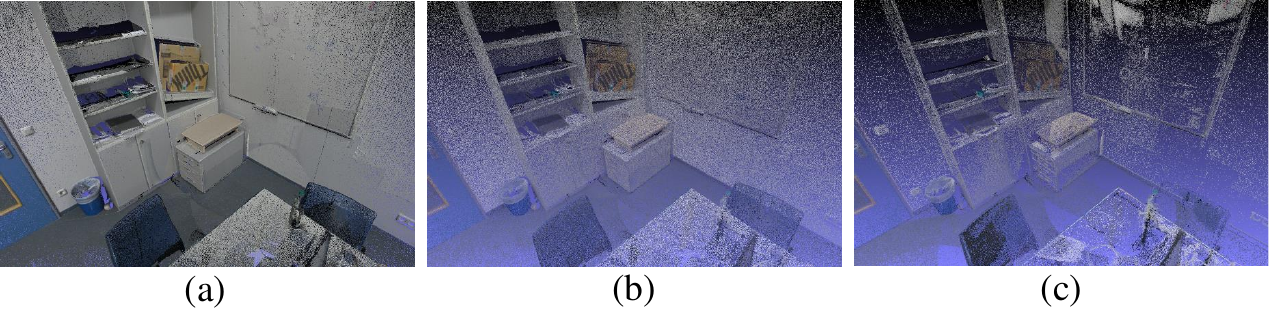}
	\caption{(a) Raw scanned point cloud from the ScanNet++ dataset. (b) Randomly downsampled point cloud. (c) Geometry-texture-aware point cloud allocation under a fixed budget.}
	\label{fig:geotex}
\end{figure}

\subsection{Geometry-Conditioned Gaussian Allocation}
\label{sec:allocation}

The allocation stage aims to initialize a fixed number of Gaussian primitives from dense LiDAR point clouds while preserving critical geometric and appearance information. Let $\mathcal{P} = \{p_i\}_{i=1}^{N}$ denote the registered LiDAR point cloud, where $N$ can exceed tens of millions. Directly initializing Gaussian primitives from all points is computationally prohibitive. Therefore, the objective is to select a subset $\mathcal{P}' \subset \mathcal{P}$ with $|\mathcal{P}'| = M \ll N$ that maximizes information retention under a fixed Gaussian budget $M$.

\paragraph{Geometry and Appearance Complexity Estimation}
The local information content of each point is characterized using both geometric and appearance cues. For each point $p_i$, a local neighborhood $\mathcal{N}_i^k$ is constructed via $k$-nearest neighbors, and geometric complexity is estimated through curvature analysis. Specifically, the covariance matrix is computed as
\begin{equation}
\mathbf{C}_i = \frac{1}{k-1} \sum_{p_j \in \mathcal{N}_i^k} (p_j - \mu_i)(p_j - \mu_i)^{T},
\end{equation}
where $\mu_i$ denotes the centroid of $\mathcal{N}_i^k$. Let $\lambda_1 \leq \lambda_2 \leq \lambda_3$ be the eigenvalues of $\mathbf{C}_i$. The normalized curvature is defined as
\begin{equation}
\kappa_i = \frac{\lambda_1}{\lambda_1 + \lambda_2 + \lambda_3 + \epsilon},
\end{equation}
which serves as a proxy for local surface complexity. Larger curvature values typically correspond to edges, corners, and thin structures that require denser Gaussian representation.

When color information is available, appearance complexity is further estimated based on local color variation. Let $c_j \in \mathbb{R}^3$ denote the RGB value of a neighboring point $p_j$. The texture complexity $\tau_i$ is computed as
\begin{equation}
\tau_i = \frac{1}{3k} \sum_{m=1}^{3} \sum_{p_j \in \mathcal{N}_i^k} \left(c_j^{(m)} - \bar{c}_i^{(m)}\right)^2,
\end{equation}
where $\bar{c}_i^{(m)}$ represents the mean color value of channel $m$. This term captures appearance variation in textured or color-inhomogeneous regions.

\begin{figure}[tbp]
	\centering
	\includegraphics[width=\linewidth]{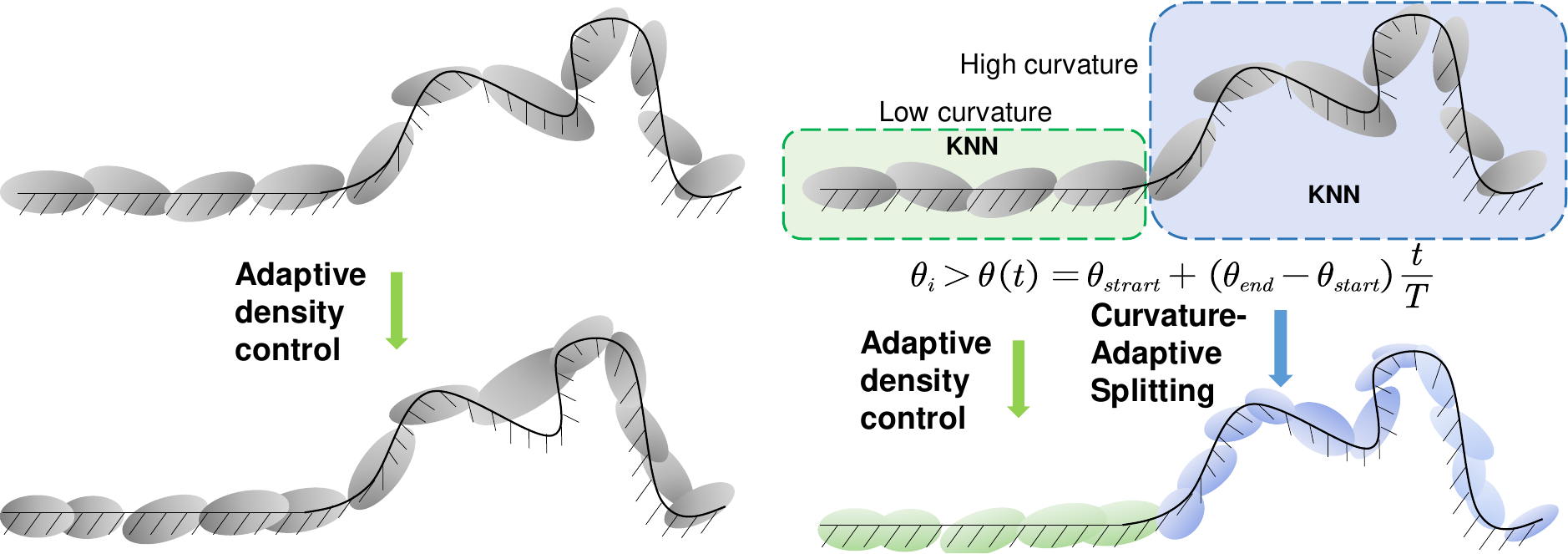}
	\caption{Curvature-adaptive splitting enables finer-grained refinement in geometrically complex regions, thereby preserving detailed structural geometry.}
	\label{fig:cur_split}
\end{figure}

\paragraph{Information-Aware Allocation under a Fixed Budget}
Both geometric and appearance complexity measures are normalized to the range $[0,1]$ and combined to define an information-aware sampling distribution. The allocation probability for each point is defined as
\begin{equation}
P_i = \frac{\alpha \hat{\kappa}_i + \beta \hat{\tau}_i}{\sum_{j=1}^{N} (\alpha \hat{\kappa}_j + \beta \hat{\tau}_j)},
\end{equation}
where $\alpha$ and $\beta$ control the relative contributions of geometric and appearance cues, with $\alpha + \beta = 1$. A total of $M$ points are then sampled from $\mathcal{P}$ according to $P_i$ to form the initial Gaussian set $\mathcal{P}'$.

\paragraph{Scalability Considerations}
To handle large-scale LiDAR point clouds, the allocation process is performed in a chunk-wise manner with CUDA-accelerated nearest neighbor queries to ensure memory efficiency and scalability. In all experiments, the Gaussian budget $M$ is fixed to a constant value, enabling fair comparisons with baseline sampling strategies and isolating the effect of geometry-conditioned allocation.

\subsection{Geometry-Aware Gaussian Refinement via Curvature-Adaptive Splitting}
\label{sec:refinement}

While geometry-conditioned allocation provides an informative initialization, structural degradation may still occur during training due to the view-dependent densification mechanism in vanilla 3DGS. Since Gaussian splitting is primarily driven by photometric gradients and opacity cues, excessive refinement may arise in large planar regions, while thin structures or sharp geometric boundaries may remain under-refined, particularly under sparse or uneven viewpoints.

To introduce explicit geometric guidance into densification, a curvature-adaptive splitting strategy is employed. The central idea is to use local surface curvature as a geometry-aware indicator of where additional representational capacity is required. Compared to view-driven heuristics, curvature provides a scene-structure prior that is less sensitive to view distribution.

\paragraph{Online Curvature Estimation on Gaussians}
During training, a curvature score $\kappa_i$ is computed for each Gaussian primitive $G_i$ based on the current spatial distribution of Gaussian centers. The curvature definition follows the same formulation introduced in Sec.~\ref{sec:allocation}, but is evaluated on the evolving Gaussian set rather than the original LiDAR points. In this manner, curvature reflects the current local geometric configuration and is updated as Gaussians move or are split during optimization.

\paragraph{Curvature-Adaptive Splitting Rule}
Gaussian splitting is modulated according to the estimated curvature. A Gaussian primitive $G_i$ is selected for splitting if its associated curvature value exceeds a dynamically scheduled threshold:
\begin{equation}
\kappa_i > \theta(t),
\end{equation}
where $\kappa_i$ is obtained from the curvature estimation described above, and $\theta(t)$ denotes a time-dependent threshold.

To ensure stable optimization and progressive refinement, a coarse-to-fine scheduling strategy is adopted for the curvature threshold:
\begin{equation}
\theta(t) = \theta_{\text{start}} + \left( \theta_{\text{end}} - \theta_{\text{start}} \right) \frac{t}{T},
\end{equation}
where $T$ represents the total number of training iterations. This schedule allows broader refinement at early stages to establish coarse structure, while progressively focusing splitting on higher-curvature regions to recover fine geometric details.

\paragraph{LiDAR-Normal Assisted Structure Preservation}
While curvature-adaptive splitting determines where refinement is required, additional orientation constraints help stabilize local surface structure during optimization. Accordingly, a normal consistency regularization based on LiDAR-derived surface normals is introduced.

For each Gaussian primitive $G_i$ with covariance $\boldsymbol{\Sigma}_i$, a local Gaussian normal $\mathbf{n}_i^{\text{gs}}$ is extracted as the eigenvector corresponding to the smallest eigenvalue of $\boldsymbol{\Sigma}_i$, representing the local surface normal implied by the anisotropic Gaussian.

Each Gaussian is associated with a LiDAR surface normal $\mathbf{n}_i^{\text{lidar}}$ via nearest-neighbor correspondence in the registered LiDAR point cloud. A normal alignment loss is defined as
\begin{equation}
\mathcal{L}_{\text{normal}}
=
\frac{1}{|\mathcal{G}|}
\sum_{i}
\left(
1 -
\left|
\mathbf{n}_i^{\text{gs}} \cdot \mathbf{n}_i^{\text{lidar}}
\right|
\right),
\end{equation}
which penalizes angular deviation between Gaussian-implied normals and LiDAR surface normals. This regularization stabilizes local surface orientation, particularly around thin structures and sharp edges, while remaining independent of the curvature-adaptive splitting criterion.

% \paragraph{Structure-Preserving Gaussian Refinement.}
% For each selected Gaussian primitive, we perform splitting by generating multiple offspring Gaussians with perturbed positions and reduced covariance scales, following the standard 3DGS splitting procedure. Unlike view-dependent densification strategies that indiscriminately increase Gaussian density, the proposed curvature-adaptive refinement redistributes representational capacity toward geometrically complex regions. As a result, planar surfaces naturally converge with fewer primitives, while edges and thin structures are progressively refined with finer-grained Gaussian representations. 

% This geometry-aware refinement process mitigates structural blurring commonly observed under sparse viewpoints and complements the geometry-conditioned allocation stage by dynamically adjusting Gaussian distribution during training based on evolving geometric cues. Together, the two stages enable a unified geometry-conditioned optimization process that preserves structural details while maintaining efficient utilization of Gaussian primitives.

\begin{figure}[tbp]
	\centering
	\includegraphics[width=\linewidth]{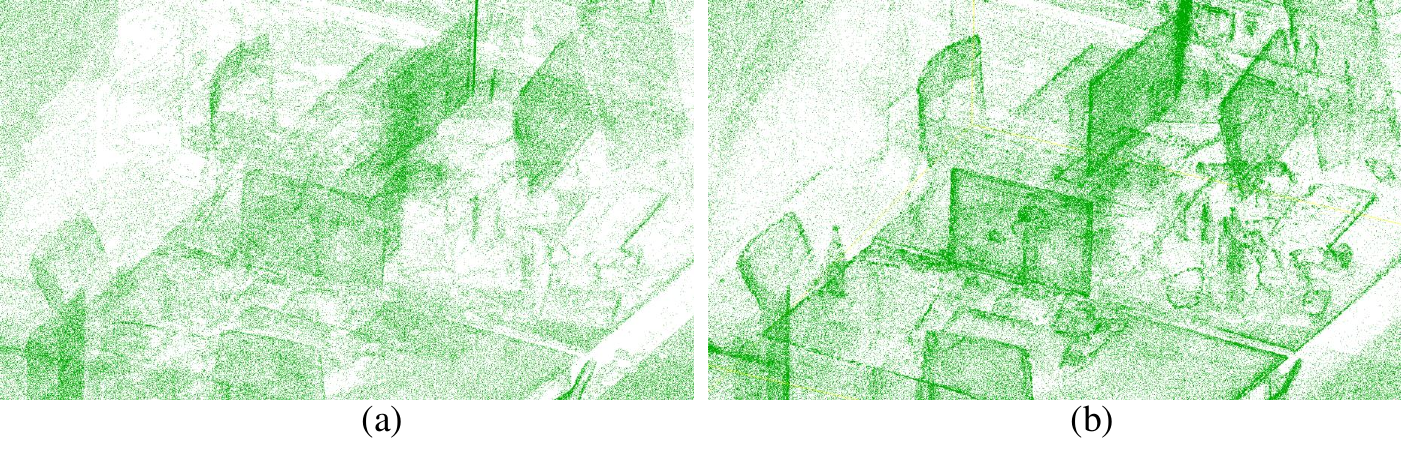}
	\caption{Initialized with randomly sampled LiDAR point clouds. (a) Point cloud distribution of the original 3DGS. (b) Point cloud distribution after applying the geometry-aware splitting strategy.}
	\label{fig:cur_points}
\end{figure}

\subsection{Confidence-aware Metric LiDAR Depth Regularization}
\label{sec:metric_constraint}

While geometry-conditioned allocation and refinement improve the structural fidelity of Gaussian representations, photometric supervision alone remains insufficient to enforce absolute scale consistency. In particular, under sparse viewpoints or texture-degenerate regions, image-based losses may admit multiple geometrically plausible but metrically inconsistent solutions, resulting in scale drift and floating artifacts. This limitation is intrinsic to vision-only reconstruction and persists even when Gaussian distributions are well allocated and refined.

\begin{figure}[tbp]
	\centering
	\includegraphics[width=\linewidth]{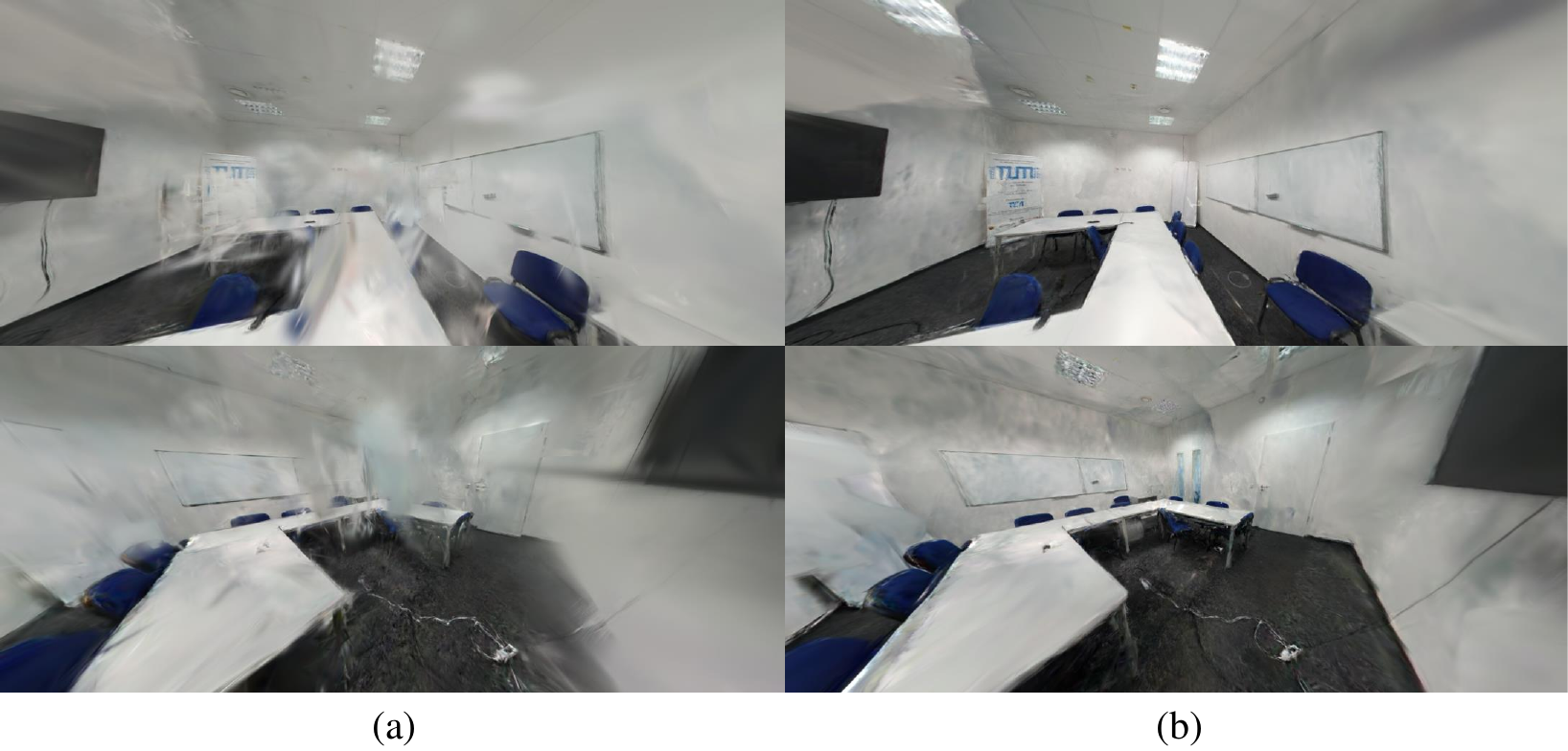}
	\caption{Sparse-condition experimental results on the ScanNet++ dataset. (a) Without LiDAR-based metric-scale depth constraints. (b) With LiDAR-based metric-scale depth constraints.}
	\label{fig:depth}
\end{figure}

To address this issue, a metric-scale constraint is introduced by incorporating LiDAR-derived depth measurements into the Gaussian optimization process. Unlike relative depth cues inferred from images, LiDAR provides absolute geometric measurements in real-world units, offering a principled mechanism to anchor Gaussian primitives to metric space.

\paragraph{Metric Depth Extraction from Registered LiDAR}
Given the globally registered LiDAR point cloud $\mathcal{P}$ and calibrated camera poses, LiDAR points are projected into each camera view to obtain sparse but metrically accurate depth observations. Let $D_{\text{LiDAR}}(p)$ denote the LiDAR depth value at pixel $p$ after projection and visibility filtering. These depth measurements serve as metric supervision signals during training.

\paragraph{Unbiased Depth Rendering with Gaussian Primitives}
To compare LiDAR depth with the rendered geometry, an unbiased depth rendering formulation based on planar Gaussian primitives is adopted, following prior work on geometry-aware Gaussian splatting. For each Gaussian primitive $G_i$, a local surface plane is defined by its mean $\boldsymbol{\mu}_i$ and covariance $\boldsymbol{\Sigma}_i$. The plane normal $\mathbf{n}_i$ and plane-to-camera distance $\mathcal{D}_i$ are derived from the eigen-decomposition of $\boldsymbol{\Sigma}_i$.

\begin{figure*}[t!]
	\centering
	\includegraphics[width=\linewidth]{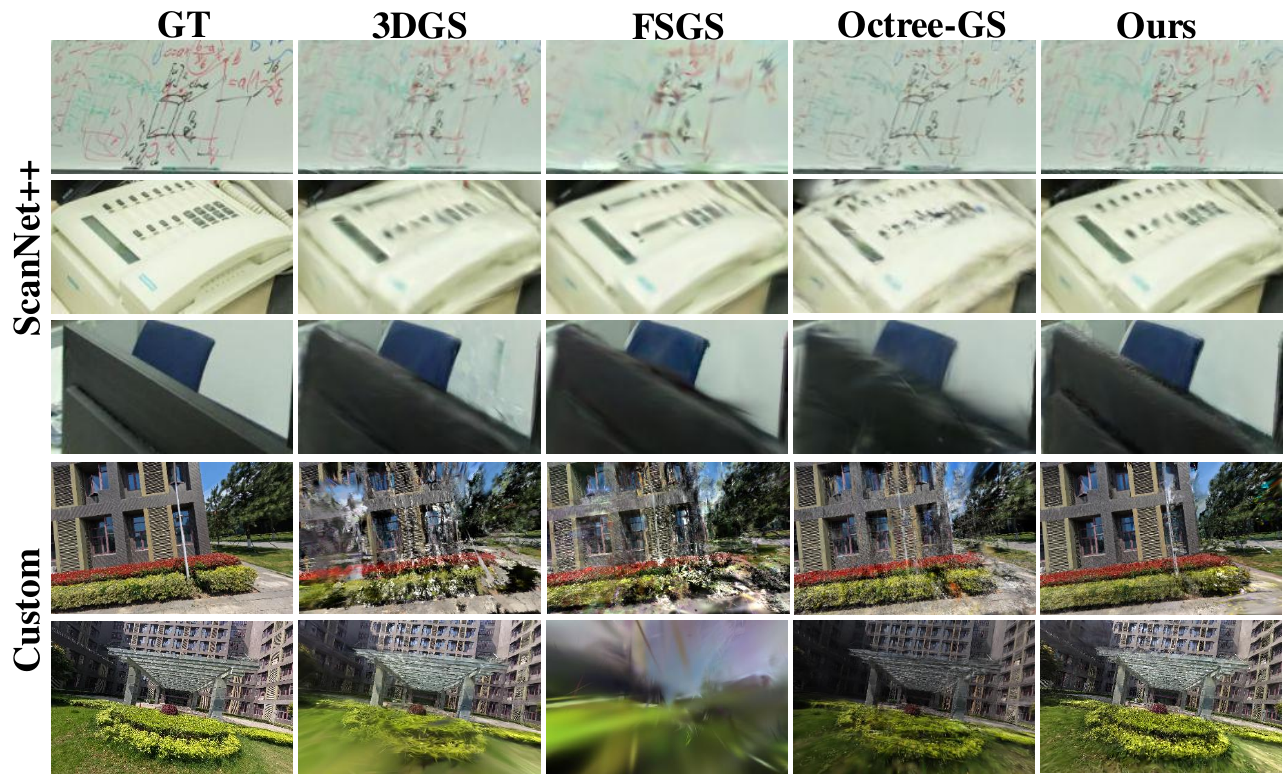}
	\caption{Qualitative comparison on ScanNet++ dataset and custom dataset.}
	\label{fig:compare}
\end{figure*}

Given a camera ray corresponding to pixel $p$, the rendered depth $\hat{D}(p)$ is computed as
\begin{equation}
\hat{D}(p) =
\frac{
\sum_{i \in \mathcal{V}(p)} \alpha_i \, \mathcal{D}_i
}{
\sum_{i \in \mathcal{V}(p)} \alpha_i \, (\mathbf{n}_i^{\top} \mathbf{K}^{-1} \tilde{p})
},
\end{equation}
where $\mathcal{V}(p)$ denotes the set of Gaussian primitives contributing to pixel $p$, $\alpha_i$ is the opacity of $G_i$, $\mathbf{K}$ is the camera intrinsic matrix, and $\tilde{p}$ represents the homogeneous image coordinate. This formulation avoids depth bias caused by opacity-weighted accumulation and ensures that rendered depth values lie on the reconstructed surface.

\paragraph{Confidence-Aware Metric Depth Regularization}
Although LiDAR depth measurements provide metric-scale accuracy, they may exhibit noise, sparsity, and misalignment near depth discontinuities, thin structures, and occlusion boundaries. Enforcing hard depth constraints uniformly across all regions may destabilize optimization and adversely affect photometric consistency. To address this issue, a confidence-aware weighting scheme is adopted to modulate the influence of LiDAR supervision based on local image structure.

A per-pixel confidence weight $w(p)$ is computed from image gradients, which serve as a proxy for depth reliability:
\begin{equation}
w(p) = 1 - \frac{\left| \nabla^2 I(p) \right|}{\max_{q \in \mathcal{U}} \left| \nabla^2 I(q) \right|},
\end{equation}
where $\nabla^2 I(p)$ denotes the image Laplacian at pixel $p$, and $\mathcal{U}$ represents the set of pixels with valid LiDAR depth. This formulation assigns lower confidence to high-frequency regions, such as edges and occlusions, and higher confidence to geometrically stable areas.

The metric depth loss is defined as
\begin{equation}
\mathcal{L}_{\text{depth}} =
\frac{1}{|\mathcal{U}|}
\sum_{p \in \mathcal{U}}
w(p) \, \left| D_{\text{LiDAR}}(p) - \hat{D}(p) \right|.
\end{equation}

The overall training objective combines photometric reconstruction losses with metric-scale regularization:
\begin{equation}
\mathcal{L} =
\lambda_{\text{rgb}} \mathcal{L}_{\text{rgb}} +
\lambda_{\text{ssim}} \mathcal{L}_{\text{ssim}} +
\lambda_{\text{depth}} \mathcal{L}_{\text{depth}},
\end{equation}
where $\lambda_{\text{depth}}$ controls the contribution of metric depth supervision. A moderate weighting balances metric anchoring with photometric optimization, preventing over-constraining effects during training.

% \begin{table*}[htbp]
%     \centering
%     \caption{Quantitative comparison of rendering quality between indoor scenes on ScanNet++ and outdoor scenes on custom dataset.}
%     \label{tab:compare}
%     \renewcommand{\arraystretch}{1.2}
%     \setlength{\tabcolsep}{8pt}
%     \begin{tabular}{c|ccc|ccc}
%         \toprule
%         Dataset & \multicolumn{3}{c|}{ScanNet++} & \multicolumn{3}{c}{Custom} \\
%         \begin{tabular}{c|c} Method & Metrics \end{tabular}  & PSNR \(\uparrow\) & SSIM \(\uparrow\) & LPIPS \(\downarrow\) & PSNR \(\uparrow\) & SSIM \(\uparrow\) & LPIPS \(\downarrow\) \\
%         \midrule
%         \textbf{3DGS} & 21.6420 & \cellcolor{tabthird}0.8271 & \cellcolor{tabsecond}0.2888 & \cellcolor{tabsecond}13.9350 & \cellcolor{tabsecond}0.2862 & \cellcolor{tabthird}0.5670 \\
%         \textbf{FSGS} & \cellcolor{tabsecond}22.0859 & \cellcolor{tabfirst}0.8381 & \cellcolor{tabthird}0.2920 & 12.2362 & 0.2566 & 0.6396 \\
%         \textbf{Octree-GS} & \cellcolor{tabthird}21.6805 & 0.8244 & 0.2928 & \cellcolor{tabthird}13.1322 & \cellcolor{tabthird}0.2784 & \cellcolor{tabsecond}0.5343 \\
%         \textbf{Ours} & \cellcolor{tabfirst}22.2463 & \cellcolor{tabsecond}0.8364 & \cellcolor{tabfirst}0.2761 & \cellcolor{tabfirst}15.2345 &\cellcolor{tabfirst}0.3714 &\cellcolor{tabfirst}0.4579 \\
%         \bottomrule
%     \end{tabular}
% \end{table*}

\begin{table*}[htbp]
    \centering
    \caption{Quantitative comparison of rendering quality and computational cost on ScanNet++ and custom datasets.}
    \label{tab:compare}
    \renewcommand{\arraystretch}{1.2}
    \setlength{\tabcolsep}{8pt}
    \begin{tabular}{c|ccc|ccc|cc}
        \toprule
        \multirow{2}{*}{Method} 
        & \multicolumn{3}{c|}{ScanNet++} 
        & \multicolumn{3}{c|}{Custom}
        & \multicolumn{2}{c}{Cost} \\
        
        & PSNR $\uparrow$ & SSIM $\uparrow$ & LPIPS $\downarrow$
        & PSNR $\uparrow$ & SSIM $\uparrow$ & LPIPS $\downarrow$
        & Time & GPU Mem. \\
        \midrule
        
        \textbf{3DGS} 
        & 21.6420 & \cellcolor{tabthird}0.8271 & \cellcolor{tabsecond}0.2888
        & \cellcolor{tabsecond}13.9350 & \cellcolor{tabsecond}0.2862 & \cellcolor{tabthird}0.5670
        & 48 min & 5.0 GB \\
        
        \textbf{FSGS} 
        & \cellcolor{tabsecond}22.0859 & \cellcolor{tabfirst}0.8381 & \cellcolor{tabthird}0.2920
        & 12.2362 & 0.2566 & 0.6396
        & 95 min & 7.8 GB \\
        
        \textbf{Octree-GS} 
        & \cellcolor{tabthird}21.6805 & 0.8244 & 0.2928
        & \cellcolor{tabthird}13.1322 & \cellcolor{tabthird}0.2784 & \cellcolor{tabsecond}0.5343
        & 41 min & 7.2 GB \\
        
        \textbf{Ours} 
        & \cellcolor{tabfirst}22.2463 & \cellcolor{tabsecond}0.8364 & \cellcolor{tabfirst}0.2761
        & \cellcolor{tabfirst}15.2345 & \cellcolor{tabfirst}0.3714 & \cellcolor{tabfirst}0.4579
        & 15 + 43 min & 8.6 GB \\
        
        \bottomrule
    \end{tabular}
\end{table*}

\section{Experiments And Results}

\subsection{Experimental Equipment}

% We employ a custom-developed Mobile LiDAR Scanning (MLS) system to achieve synchronized acquisition of color and geometric information of indoor scenes. The system is equipped with two PandarXT16 LiDAR sensors, an Xsens MTi-630 Inertial Measurement Unit (IMU), and four GoPro Hero 10 cameras, which collectively enable comprehensive indoor scene recording.

% As illustrated in Fig.~\ref{fig:teaser}, one LiDAR sensor is mounted horizontally on the top, while the other is vertically positioned at the front. The IMU is installed directly beneath the top-mounted LiDAR, and the four GoPro cameras are symmetrically arranged in a square formation around the IMU. The two LiDAR sensors efficiently capture high-density 3D point clouds, facilitating precise indoor scene geometry acquisition with a maximum detection range of 100 meters, making them well-suited for various scene data collection tasks. The four GoPro cameras form an imaging unit, with each capable of capturing 5K-resolution images. A single synchronized shutter operation enables the acquisition of a full 360° image dataset, ensuring comprehensive and rapid scene capture. The IMU provides acceleration data for motion estimation, contributing to improved pose estimation accuracy. This integrated sensor configuration allows the MLS system to efficiently acquire and fuse multimodal data, offering robust support for detailed scene reconstruction and analysis.

As illustrated in Fig.~\ref{fig:teaser}, a mobile LiDAR scanning (MLS) system equipped with synchronized LiDAR, multi-camera, and IMU sensors is used to acquire metric-scale geometry in textureless and geometrically degenerate environments. The system enables accurate point cloud registration and multi-view image capture for the evaluation of LiDAR-guided Gaussian Splatting.

\subsection{Experimental Setup}

% \begin{figure}[t]
% 	\centering
% 	\includegraphics[width=\linewidth]{images/custom scence.pdf}
% 	\caption{Camera poses of custom dataset scenes and registered scene point clouds.}
% 	\label{fig:custom scence}
% \end{figure}

ScanNet++ provides dense indoor LiDAR scans with extensive view coverage, while our self-collected dataset intentionally adopts sparse capture to reflect practical engineering constraints, enabling evaluation under more realistic acquisition conditions.

Our implementation was built upon the official codebase of 3DGS, and incorporated the unbiased depth rendering from PGSR. Our training strategy and hyperparameters are also generally consistent with 3DGS. We implemented our method using PyTorch and trained all models on a single NVIDIA RTX 4080 SUPER GPU with 16GB memory, and 128GB RAM.

In the geometry-texture-aware adaptive sampling module, the k-nearest neighbor parameter is set to 64, with the weighting coefficients $\alpha$ and $\beta$ both assigned a value of 0.5, The target number of points for the downsampled point cloud is set to 3,000,000. In the curvature adaptive splitting strategy, $\theta_{\text{start}}$ is 0.1, $\theta_{\text{end}}$ is 0.3. In the LiDAR depth regularization, set $\lambda _{depth}$ to 1, $\lambda _{rgb}$ and $\lambda _{ssim}$ are same to 3DGS.

\subsection{Comparative Study}

We compare our method with representative 3D Gaussian Splatting approaches, including the original 3DGS, FSGS, and Octree-GS. FSGS is designed for sparse-view reconstruction through proximity-guided Gaussian unpooling, while Octree-GS adopts a level-of-detail hierarchy to improve scalability and efficiency on large-scale scenes. These methods represent complementary directions in sparse-view handling and efficiency-oriented Gaussian representations, enabling a comprehensive evaluation of our approach in terms of geometric fidelity and reconstruction quality.

\begin{table}[t]
    \centering
    \renewcommand{\arraystretch}{1.2}
    \setlength{\tabcolsep}{12pt}
    \caption{Ablation experiments on scannet++ dataset.}
    \label{tab:ablation}
    \begin{tabular}{l|ccc}
        \hline
        Model setting & PSNR$\uparrow$ & SSIM$\uparrow$ & LPIPS$\downarrow$ \\
        \hline
        None         & 21.6420 & 0.8271 & 0.2888 \\
        w/ sampling & 21.8576 & 0.8298 & 0.2903 \\
        w/ depth    & \cellcolor{tabthird}22.0951 & \cellcolor{tabsecond}0.8338 & \cellcolor{tabsecond}0.2794 \\
        w/ cur split & \cellcolor{tabsecond}22.1136 & \cellcolor{tabthird}0.8325 & \cellcolor{tabthird}0.2796 \\
        Full        & \cellcolor{tabfirst}22.2463 & \cellcolor{tabfirst}0.8364 & \cellcolor{tabfirst}0.2761 \\
        \hline
    \end{tabular}
\end{table}

As shown in Tab.~\ref{tab:compare}, the proposed method achieves competitive performance on the ScanNet++ dataset, where dense multi-view coverage reduces the performance gap between different approaches. This dataset primarily reflects well-observed indoor environments and does not fully capture the challenges encountered in real-world sparse-view settings.

To evaluate robustness under more realistic conditions, experiments are further conducted on a self-collected dataset captured with a mobile LiDAR system. As reported in Tab.~\ref{tab:compare} and illustrated in Fig.~\ref{fig:compare}, the proposed method exhibits improved structural consistency and reduced floating artifacts, particularly in outdoor and sparse-view scenarios.

Since all methods operate on dense LiDAR initialization rather than sparse SfM point clouds, the computational cost is higher than that of sparse initialization-based pipelines. Training time and peak GPU memory usage are reported in Table~\ref{tab:compare}. The runtime of the proposed method includes an offline depth preprocessing stage (15 minutes) and subsequent 3D Gaussian Splatting training (43 minutes). While neighborhood-based geometric analysis introduces additional overhead, its impact remains moderate relative to the overall training cost, and the resulting gains in geometric fidelity and stability justify this design choice.

\begin{figure}[tbp]
	\centering
	\includegraphics[width=\linewidth]{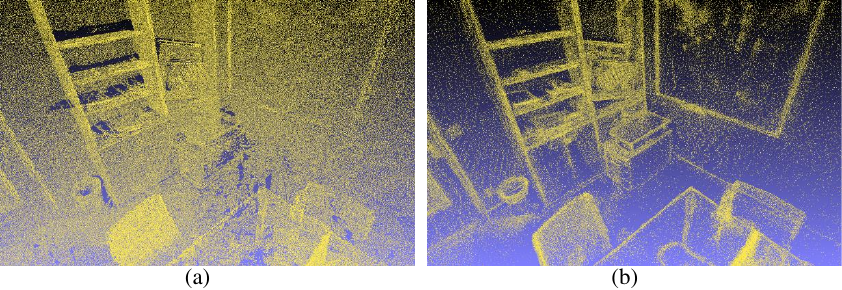}
    \caption{
    Gaussian distributions under different refinement strategies.
    (a) Vanilla 3DGS densification, where Gaussians are over-accumulated in planar regions.
    (b) Curvature-adaptive splitting, which allocates more Gaussians to high-curvature structures and reduces redundancy on planar surfaces.
    }
	\label{fig:ablation result}
\end{figure}

\subsection{Ablation Study}

Although all methods are trained under the same Gaussian budget, the initial allocation of Gaussian primitives has a significant impact on subsequent optimization. When using random or uniform downsampling, Gaussian primitives are evenly distributed across the scene, leading to a large proportion being allocated to planar regions. In contrast, geometry-texture-aware allocation concentrates Gaussian capacity in structurally complex regions, resulting in a more informative initialization under the same budget.

To isolate the effect of individual components, we conduct ablation experiments using LiDAR-registered point clouds for initialization, with both sampling strategies producing the same post-sampling point count of three million. As shown in Tab.~\ref{tab:ablation}, adding LiDAR-based metric depth supervision consistently improves rendering quality, demonstrating more stable reconstruction under the same Gaussian budget, particularly by suppressing depth drift and floating artifacts that arise under image-only supervision.

We further observe limitations in the view-dependent densification mechanism of vanilla 3DGS. Gaussian splitting is primarily driven by photometric gradients and opacity cues, which are highly sensitive to view distribution. Under sparse or uneven viewpoints, this behavior often results in excessive splitting in planar regions, where small photometric inconsistencies accumulate, while high-curvature structures such as edges and thin components remain under-refined.

As illustrated in Fig.~\ref{fig:ablation result}(a), vanilla 3DGS produces dense Gaussian clusters on planar surfaces with blurred structural boundaries. By introducing curvature-adaptive splitting, refinement is explicitly conditioned on local surface geometry. As shown in Fig.~\ref{fig:ablation result}(b), Gaussian density is progressively concentrated around structurally complex regions, while planar areas converge with fewer primitives. This geometry-aware refinement leads to sharper structural boundaries and avoids inefficient over-densification driven purely by view-dependent signals.

\section{CONCLUSIONS AND FUTURE WORK}

This paper presents a LiDAR-centric 3D Gaussian Splatting framework that alleviates key limitations of SfM-based initialization in metric-critical reconstruction scenarios. By integrating geometry-conditioned allocation, curvature-adaptive refinement, and metric-scale depth regularization, the proposed method improves geometric fidelity and scale consistency while preserving the real-time rendering advantages of 3DGS. Experiments on ScanNet++ and a custom dataset demonstrate more stable structure reconstruction and reduced artifacts under challenging observation conditions.

Nevertheless, a gap remains between commonly used benchmarks and real-world deployment. Public datasets are often captured with high-precision terrestrial laser scanners, whereas practical applications typically rely on mobile LiDAR systems with different noise characteristics, point density distributions, and motion effects. Future work will focus on extending the proposed framework to diverse mobile LiDAR settings and developing evaluation protocols that better reflect real-world operating conditions.

% \section*{Acknowledgment}

% The preferred spelling of the word ``acknowledgment'' in America is without 
% an ``e'' after the ``g''. Avoid the stilted expression ``one of us (R. B. 
% G.) thanks $\ldots$''. Instead, try ``R. B. G. thanks$\ldots$''. Put sponsor 
% acknowledgments in the unnumbered footnote on the first page.

% \section*{References}

% Please number citations consecutively within brackets \cite{b1}. The 
% sentence punctuation follows the bracket \cite{b2}. Refer simply to the reference 
% number, as in \cite{b3}---do not use ``Ref. \cite{b3}'' or ``reference \cite{b3}'' except at 
% the beginning of a sentence: ``Reference \cite{b3} was the first $\ldots$''

% Number footnotes separately in superscripts. Place the actual footnote at 
% the bottom of the column in which it was cited. Do not put footnotes in the 
% abstract or reference list. Use letters for table footnotes.

% Unless there are six authors or more give all authors' names; do not use 
% ``et al.''. Papers that have not been published, even if they have been 
% submitted for publication, should be cited as ``unpublished'' \cite{b4}. Papers 
% that have been accepted for publication should be cited as ``in press'' \cite{b5}. 
% Capitalize only the first word in a paper title, except for proper nouns and 
% element symbols.

% For papers published in translation journals, please give the English 
% citation first, followed by the original foreign-language citation \cite{b6}.

\bibliographystyle{IEEEtran}
\bibliography{IEEEabrv,reference}

\end{document}